\pdfoutput=1

\documentclass[aip,jcp,reprint,floatfix]{revtex4-2}

\usepackage{amsmath}
\usepackage{graphicx}
\usepackage{hyperref}
\usepackage{orcidlink}

\begin{document}

\title{Adsorbate Dissociation Due to Heteromolecular Electronic Energy Transfer from Fluorobenzene Thin Films}
\author{E.T. Jensen\,\orcidlink{0000-0001-8030-4204}}
\email[email:]{ejensen@unbc.ca}
\affiliation{Department of Physics \\University of Northern British Columbia, 3333 University Way, Prince George B.C. Canada V2N 4Z9}


\date{February 13, 2024}

\begin{abstract}
Study of the near-UV photodissociation dynamics for monolayer (ML) quantities of CH$_3$I on thin films of a series of fluorobenzenes and benzene (1--25ML) grown on a Cu(100) substrate finds that in addition to gas-phase-like neutral photodissociation, CH$_3$I dissociation can be enhanced via photoabsorption in several of the thin films studied. Distinct CH$_3$ photofragment kinetic energy distributions are found for CH$_3$I photodissociation on C$_6$H$_5$F, 1,4-C$_6$H$_4$F$_2$ and C$_6$H$_6$ thin films, and distinguished from neutral photodissociation pathways using polarized incident light. The effective photodissociation cross section for CH$_3$I on these thin films is increased as compared to that for the higher F-count fluorobenzene thin films due to the additional photodissociation pathway available. Quenching by the metal substrate of the photoexcitation via this new pathway suggests a significantly longer timescale for excitation than that of neutral CH$_3$I photodissociation. The observations support a mechanism in which neutral photoexcitation in the thin film (i.e. an exciton) is transported to the interface with CH$_3$I, and transferring the electronic excitation to the CH$_3$I which then dissociates. The unimodal CH$_3$ photofragment distribution and observed kinetic energies on the fluorobenzene thin films suggest that the dissociation occurs via the $^3Q_1$ excited state of CH$_3$I.
\end{abstract}

\maketitle

\section{Introduction}
Electronic energy transfer (EET) in molecular systems is a phenomenon found in a variety of contexts. For example, gas-phase photosensitization resulting in enhanced UV photodissociation of CH$_3$I in mixtures with benzene vapour was noted over 70 years ago\cite{dubois:1951}. There are many heteromolecular systems where the role of EET has been studied in detail via fluorescence (e.g. p-difluorobenzene:p-xylene\cite{lahmani:1990} and benzene:biacetyl\cite{bigman:1994} dimers) in molecular beam experiments. In condensed-phase systems of aromatic molecules the language of solid-state physics is used to describe the transport of localized electron-hole excitations in terms of Frenkel excitons (electron-hole on the same molecular site) or charge-transfer (CT) excitons when the electron-hole pair are separated on nearest-neighbour sites\cite{bardeen:2014}. In the astrophysical context, non-thermal photodesorption of H$_2$O and benzene molecules from mixed benzene:H$_2$O ices due to near-UV absorption by the benzene has been demonstrated\cite{thrower:2008, thrower:2010}. The present study expands upon an earlier work\cite{jensen:2021} in which an EET process was identified by the dynamics seen with time-of-flight spectroscopy in which contribution to the photodissociation of CH$_3$I adsorbed on benzene thin films was observed, but absent for CH$_3$I on hexafluorobenzene thin films. Here we study CH$_3$ photofragment kinetic energies from the UV photodissociation of CH$_3$I adsorbed on series of different fluorobenzene thin films and report systematic changes in the observed photodissociation dynamics.

\subsection{Near-UV Photoabsorption by Benzene and Fluorobenzenes}
The near-UV photoabsorption of condensed C$_6$H$_6$ films at 90K has been studied in detail\cite{dawes:2017,stubbing:2020}. In the region of interest for this work, the vibronic states within the $\pi$--$\pi$* transition to the $^1\!B_{2u}$ ($S_1$) electronic state are redshifted relative to the gas-phase, with the $S_1$ band origin at 4.69eV (264nm) and $6^1_01^2_0$ absorption observed at 248.85nm, close to the excimer laser wavelength used in this work. Broad spectral absorption to the triplet $^3\!E_{1u}$ ($T_2$, $\sim$225--269nm) state was also observed in addition to the $^3\!B_{1u}$ ($T_1$, $\sim$263--310nm) state. The vibronic structure of the triplet state excitations has also been detailed for condensed-phase C$_6$H$_6$ in thin films using inelastic electron scattering\cite{swiderek:1996}.

Near-UV photoabsorption for gas-phase fluoro-substituted benzenes are similar to that of C$_6$H$_6$, with the $S_1$ absorptions generally red-shifted relative to benzene and with larger cross sections\cite{philis:1981, phillips:1972}. There are few spectroscopic studies reported for condensed-phase fluorobenzenes-- a study of crystalline monofluorobenzene found the $S_1$ band origin at 4.691eV (264.3nm) with excitonic absorption features and the vibronic series extending to higher energies including the 248nm wavelength used in the present work\cite{pierre:1977}. Inelastic electron scattering has been used in studying the vibronic structure of the $T_1$ states for monofluorobenzene (mFBz) and p-difluorobenzene (dFBz) in the energy region around 3.6eV (340nm)\cite{swiderek:1997}, which lie below the energy region of interest in this work. It would be expected that the excited states of the condensed fluoro-substituted benzenes will broadly follow that of their gas-phase counterparts and have perturbation similar to that observed for benzene. We expect that the dipole-allowed absorption to the $S_1$ vibronic states will also have an overlapping $T_2$ state in the region surrounding the 248nm wavelength used in the present work\cite{phillips:1972}.

\subsection{Near-UV Photodissociation of Methyl Iodide}
\label{CH3X_pdissn}
Photodissociation of gas-phase CH$_3$I in the near-UV region occurs via the `A-band', a set of $n\rightarrow\sigma^*$ transitions observed as three overlapping states between $\sim$4.2eV and 5.6eV ($^3Q_1$, $^3Q_0$ and $^1Q_1$ in order of increasing energy) above the ground state in the Franck-Condon region\cite{Eppink:1998}. At the 248nm wavelength ($h\nu$=4.99eV) used in the present work, the $X\rightarrow {^3Q_0}$ excitation dominates gas-phase photoabsorption and the $X\rightarrow {^1Q_1}$ is a minor channel. The subsequent dissociation can proceed via two principal pathways:
\begin{equation} \label{Equ_1}
\begin{split}
CH_3I + h\nu & \rightarrow CH_3 + I({^2\!P_{3/2}}) \textrm{ \{ground state I\}} \\
& \rightarrow CH_3 + I^*(^2\!P_{1/2}) \textrm{ \{excited I\}}
\end{split}
\end{equation}

The energy difference between ground state $I$ and excited $I^*$ is 0.943eV, leading to significant differences in the translational energies imparted to the fragments and which can be resolved in our time-of-flight measurements. There are also vibrational and rotational energy partitioning differences for the CH$_3$ photofragments along the two pathways. Another significant factor for this system is that the $X\! -\! {^3Q_0}$ excitation is a parallel transition (requiring a component of the incident $\vec{E}$-field along the C--I bond axis), while the ${^1Q_1}$ and $^3Q_1$ excitations are perpendicular. This polarization dependence for optical absorption allows utilizing polarization and molecular orientation to aid in understanding the photodissociation dynamics at 248nm\cite{Jensen:2005}. The $^3Q_0$ state correlates to the I$^*$ outcome in Equ.~{\ref{Equ_1}}, but a curve-crossing with the $^1Q_1$ state (which correlates to the I pathway) during dissociation enables non-adiabatic transitions that result in both pathways being observed subsequent to an initial $^3Q_0$ excitation\cite{Eppink:1998}.

\subsection{Energetics of Stimulated Dissociation}
The dissociation of a CH$_3$I molecule in free space requires momentum and energy conservation, which determines how the excess kinetic energy is partitioned between the CH$_3$ fragment and the halogen atom. For neutral photodissociation, a starting point for rationalizing the CH$_3$ photofragment kinetic energy is:

\begin{eqnarray}
T_{CH_3}
& =\frac{m(I)}{m(CH_3I)} \{ &h\nu - D_0(C-I) -  E_{int}(I) \nonumber \\
& & - E_{int}(CH_3) \}
\label{Equ_2}
\end{eqnarray}

%

where $m()$ is the mass of the particular species, $h\nu$ is the photon energy, $D_0$ is the bond energy, and $E_{int}(I)$ allows for the possible electronic excitation of the departing halogen atom and $E_{int}(CH_3)$ is the internal energy (vibration and rotation) of the departing methyl fragment.

For surface adsorbate systems the parent molecule is not in free space, but embedded at or near the vacuum interface of the system being studied. The departing CH$_3$ fragments must also escape the surface attractive forces, so the kinetic energies will be reduced by an amount on the order of 0.1eV\cite{altaleb:2017}. Further, it is known from prior work in gas-phase cluster and surface photochemistry that the observed fragment kinetic energy distributions can be altered by chemical or post-dissociation interactions, however Equ.~{\ref{Equ_2}} provide a basis to begin rationalization of the observed kinetic energy distributions.

\subsection{Adsorption and Valence Band Structure of Fluorobenzenes on the Cu(100) Surface}
The adsorption behaviour of monolayer and thin films of C$_6$H$_6$ and C$_6$F$_6$ on coinage metal surfaces has been studied and described in detail in the literature. Studies of C$_6$H$_6$ on copper surfaces has revealed differences in structure for the first layer that depend on the particular surface (material and crystal face), most commonly with molecules lying flat for the first layer and tilting upright in the second layer\cite{lee:2006,xi:1994}, behaviour ascribed to the quadrupole moment for the molecule and a propensity for a T-motif in the solid\cite{xi:1994}. For thicker benzene films a herringbone structure is believed to occur\cite{jakob:1996}, similar to the crystal structure of the bulk solid. By contrast, C$_6$F$_6$ is believed to adsorb in flat-lying layers due to the F-atoms forming halogen bonds with C--C bonds of neighbouring molecules\cite{Vijayalakshmi:2006ka,zhao:2014}. In contrast to the case for C$_6$H$_6$ and C$_6$F$_6$, there are relatively few surface science studies in the literature that have characterized the adsorption of other fluorobenzenes on coinage metal surfaces. The various factors that affect the thin film structure, such as the surface corrugation and lattice spacing as well as the intermolecular interactions such as electrostatic multipole moments, hydrogen bonding or halogen bonding will be significant for the fluorobenzenes. For example, infrared spectroscopy of C$_6$H$_5$F on Cu(111) finds that the molecules lie flat until the first layer is ~2/3 complete, and then the molecules distort and tilt with the C--F bond coming out-of-plane\cite{schunke:2022}. A scanning probe microscopy study of the adsorption of 1,3,5-C$_6$H$_3$F$_3$ on the more corrugated Ag(110) surface observed a rectangular unit cell with alternate molecules rotated to facilitate halogen bonding with its neighbours\cite{han:2017}.

Although we report some data for fluorobenzene thin films of 1ML coverage, the focus of the present work is on multilayer thin films, with the majority of the data presented using 10ML thin films. As such we are interested in the bulk-like crystal structures that are likely to be formed under our experimental conditions. In the present work a monolayer of CH$_3$I is adsorbed on top of the fluorobenzene thin film, so the structure of the CH$_3$I layer will be influenced by the underlying thin film structure, primarily through the electrostatic multipole moments of the particular species on the condensed solid. Given the large number of possible fluorobenzene molecules, we restricted the present study to those with dipole moments of zero: C$_6$F$_6$ (hFBz), 1,2,4,5-C$_6$H$_2$F$_4$ (tetraFBz), 1,3,5-C$_6$H$_3$F$_3$ (triFBz), 1,4-C$_6$H$_4$F$_2$ (dFBz) and C$_6$H$_6$ (benzene); plus C$_6$HF$_5$ (pFBz) and C$_6$H$_5$F (mFBz) which possess non-zero dipole moments. The systematics of bulk crystal structures for the fluorobenzenes are described in detail in Ref. {\onlinecite{thalladi:1998}}. 

As for the studies of physical structure, previous studies of the electronic structure of these molecules in the adsorbed state has been extensive for benzene as well as for hFBz, these have been studied using one-photon and two-photon photoemission\cite{vondrak:1999,gahl:2000, dutton:2001}, inverse photoemission\cite{dudde:1990} and the subject of some theoretical electronic bandstructure calculations\cite{zhao:2014,han:2017}. There is relatively smaller publication list on the electronic structure for the intermediate fluorobenzenes\cite{miller:2015}, with work published using inelastic electron scattering for studies of excitations in dFBz and mFBz thin films\cite{swiderek:1997}.

\section{Experimental Details}
\label{expt_detail}
The experiments were performed in an ultra-high vacuum (UHV) system that has been described previously\cite{Jensen:2005}. The Cu(100) single crystal sample is 12mm in diameter and was cooled by liquid nitrogen (base temperature $\sim$90K) and heated by radiative heating from a filament up to 300K and electron bombardment heating to 920K for cleaning. Sample temperatures were monitored by a type K thermocouple spot-welded to the tungsten sample mounting wire. Sample cleanliness and order were monitored by Auger electron spectroscopy (AES) and low energy electron diffraction (LEED) measurements respectively. The crystal was prepared in UHV by cycles of Ar$^+$ ion bombardment and electron bombardment heating and annealing until the sample AES spectra indicated a clean copper substrate with the LEED patterns of a (1$\times$1) surface.

Deposition of molecules on the sample was done using a micro-capillary array directed doser\cite{Fisher:2005uw}, with the sample held normal to the doser, 25mm away. This arrangement was found to enhance the deposition by a factor of 10 compared to background dosing. The pressure in the UHV chamber was measured using uncorrected ionization gauge readings. The dosing (in Langmuirs, L) was calibrated in terms of equivalent monolayers for the different species used by Temperature Programmed Desorption (TPD) measurements as discussed in Section {\ref{TPD_section}} below. The liquids used in this work were degassed by multiple freeze-pump-thaw cycles. Vapour from CH$_3$I (Sigma-Aldrich, 99.5\%), benzene (Sigma-Aldrich, 99.8\%), mFBz (Sigma-Aldrich, 99\%), dFBz (Sigma-Aldrich, \textgreater 99\%), triFBz (TCI, \textgreater 98\%), tetraFBz (Sigma-Aldrich, \textgreater 99\%), pFBz (TCI, \textgreater 98\%) and hFBz (Sigma-Aldrich, \textgreater 99.5\%) was obtained from the liquid in a pyrex vial a few cm from the precision leak valve used to admit the room-temperature vapour to the directed doser. The deposition of the molecules was done using substrate temperatures below 100K.

Temperature programmed desorption (TPD) measurements were made by positioning the sample to face a quadrupole mass spectrometer (QMS). In early experiments, a UTI 100C QMS was used, and for later experiments it was replaced by a Stanford Research Systems RGA 200 QMS. The QMS ionizer was located $\sim$80mm away from the sample and behind an aperture that limits the ionizer line-of-sight to the central region of the sample. The sample was heated using a filament at ground potential, located a few mm behind the sample mount.

The time-of-flight (TOF) photodissociation measurements were performed using a second QMS (Extrel). Neutral products from the Cu(100) surface travel 185mm to pass through a 4mm diameter aperture to a differentially pumped region with an axial electron bombardment ionizer. The sample to ionizer distance is 203mm. Ions created in the ionizer travel through the quadrupole region and are mass selected, in the present experiments using m/q=15amu. Ion arrivals are recorded using a multichannel scaler that begins counting 50$\mu s$ prior to the initiating laser pulse, and the counts recorded from multiple laser pulses are summed. Unless otherwise indicated, the spectra shown in the present work are the result of summing data from 1000 laser pulses into 1000 1$\mu s$ time bins. In order for the ion arrival times to reflect the neutral fragment time-of-flight, they are corrected for the ion flight time (for CH$_3^+$, 17$\mu s$ at the 50eV ion energy used in the QMS). This is the leading systematic uncertainty in the recorded flight times ($\pm 1.5\mu s$) which does not affect comparisons between different TOF spectra but does lead to fixed nonlinear systematic uncertainty in the reported fragment kinetic energies $(KE\propto 1/(TOF)^2)$, which is most problematic at short flight times. The TOF spectra $N(t)$ were converted to probability distributions $P(E)$ versus CH$_3$ kinetic energy using the Jacobian transformation with an added factor $1/t$ to account for the higher ionization probability for slower neutral CH$_3$ fragments\cite{zimmermann:1995}. 

The laser pulses ($\sim$5ns duration) are produced by a small excimer laser (MPB PSX-100) operating at 20Hz. In this work KrF ($\lambda$=248nm, $h\nu$=4.99eV) laser light was used, with laser fluences on the sample of $\sim$0.8mJ/cm$^2$. The intrinsic bandwidth of the laser emission for excimer lasers is rather broad-- for a free-running KrF excimer laser the center wavelength is approximately 248.4nm (4.992eV) and has a fwhm bandwidth of $\sim$0.40nm (0.008eV). 

Linearly polarized laser light has been used exclusively in this work for the reasons described in Section {\ref{CH3X_pdissn}}. To create the polarized light, the beam is incident upon a birefringent MgF$_2$ crystal prism at the Brewster angle to separate the p- and s-polarized components. The p-polarized beam was aligned on the sample. The s-polarized light was derived from the p-polarized beam by inserting an antireflection coated zero-order half-waveplate into the beam. The laser light was collimated using a 6mm diameter aperture and was unfocused on the sample. The laser light is incident upon the sample at a fixed angle of 45$^\circ$ from the TOF mass spectrometer axis-- for example, when the Cu crystal sample is oriented to collect desorption fragments along the surface normal direction, the light is incident at 45$^\circ$.

Cross sections for the photodissociation of molecular thin films examined in this work were determined by measuring the depletion rate of the CH$_3$ photofragment yields (summed CH$_3$ photofragment counts). These ``depletion cross sections'' are obtained by recording CH$_3$ photofragment yields from photodissociation for a sequence of TOF spectra. Time-of-flight spectra are obtained using 200--400 laser pulses per scan, then repeated for 10 or more successive scans. In the systems studied here, the photofragment yields are observed to monotonically diminish as the net laser photon flux was increased, and the resulting yield vs. photon flux curves were then fit using a single exponential decay model. Reasonable fits to the data were obtained, at least in the low photon flux region. This procedure allows the possibility that other photochemical processes involving CH$_3$I removal but not seen in the TOF data might be occurring in the heterogeneous thin films. In the present work, the dissociated species was always from a dose equivalent to 1ML CH$_3$I film, while the underlying fluorobenzene thin film was varied. We used a 1ML CH$_3$I on a 10ML C$_6$F$_6$ thin film as a reference system for comparison when studying the cross sections of the other fluorobenzene thin films. The absolute cross sections measured this way have fairly large errors (we estimate $\pm50\%$), but the errors in comparative relative cross sections (i.e. vs CH$_3$I on C$_6$F$_6$) between the different systems is much lower, on the order of $\sim$10\% based on repeatability of measurements.

\section{Results and Observations}
\subsection{Temperature Programmed Desorption of Fluorobenzenes on Cu(100)}
\label{TPD_section}
In order to be able to adsorb known quantities of the various molecules used in the present work, TPD has been used to characterize the relationship between the amounts dosed (in Langmuirs, L) and the formation of a complete monolayer (1ML). After establishing the dose required to form 1ML for a particular molecule, we have assumed that for the temperature used in these experiments (T\textless 100K) the sticking coefficient for these molecules is close to unity for multilayer films, so dose amounts in multiples of the monolayer dose to produce the desired multilayer thin film. We have characterized TPD for CH$_3$I, C$_6$H$_6$ and C$_6$F$_6$ in our apparatus previously{\cite{jensen:2021}} so have performed TPD experiments for the other fluorobenzenes used in the present work. 

The TPD spectra observed for the sequence from mFBz through pFBz were similar in terms of general features to those observed for benzene\cite{jensen:2021, lee:2006, xi:1994} and hFBz\cite{jensen:2021, vondrak:1999}. Submonolayer quantities of the fluorobenzenes were observed to desorb from the Cu(100) substrate with a peak near 210K (ranging from 200K to 220K). As the dose was increased, the monolayer desorption peak broadened toward lower temperatures, characteristic of repulsive intermolecular interactions. At the completion of the first monolayer, the monolayer desorption peaks extended from $\sim$160K to just above 220K. A peak corresponding to desorption from the second layer emerged at 150-165K depending the particular fluorobenzene, and this peak shifted to slightly higher temperature as the second layer coverage was increased, with a peak in the desorption temperature 160K-170K. For mFBz, a peak near 158K emerged for doses that were compatible with growth of the third layer (where the second layer peak was at 164K), and as this peak grew it merged with the bilayer peak at higher doses. For the other fluorobenzenes in this work, we did not observe differences in desorption peaks between the second and third or subsequent layers, aside from growing peak height with dose. For the fluorobenzenes in the current study, doses corresponding to the completion of 1ML were found to be in the range 0.38L to 0.43L though with small uncertainties (5-10\%). This was used as the basis for creating the multilayer films-- doses are quoted in nominal monolayer (ML) amounts and as will be discussed in Section \ref{filmmorph}, there are reasons to believe that some of the multilayer fluorobenzene films exhibit Stranski-Krastanov growth.

\subsection{Photodissociation of CH$_3$I on Fluorobenzene Thin Films}
\begin{figure}[h]
\includegraphics[scale=0.65]{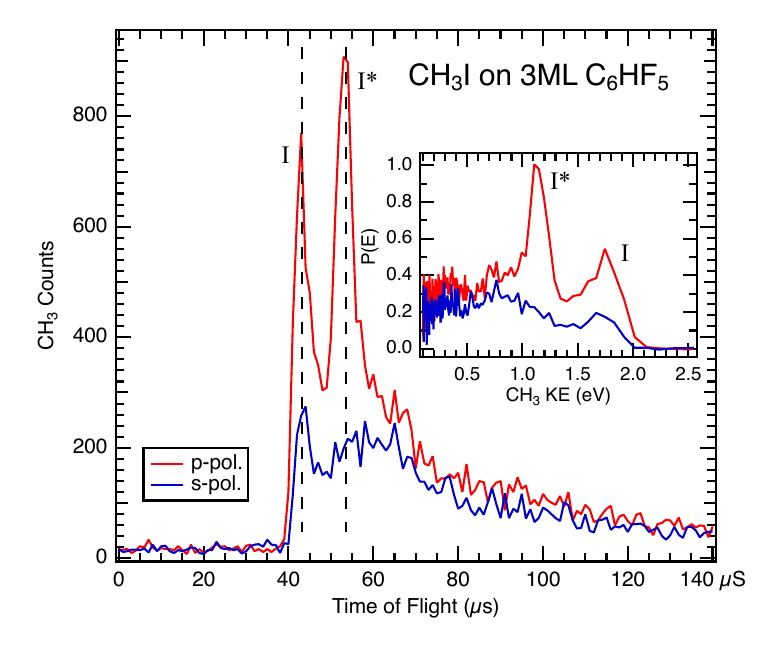}
\caption{Time-of-flight spectra for CH\(_3\) photofragments from the photodissociation of 1ML CH$_3$I adsorbed on 3ML C\(_6\)HF\(_5\) (pFBz) on Cu(100). The  red and blue traces are obtained using p- and s-polarized 248nm light respectively. The CH$_3$ photofragments are detected in the surface normal direction. The vertical dashed lines indicate the nominal flight times for the CH$_3$ photofragment peaks from the I and I* A-band photodissociation pathways as described in the text. The inset plots show the data plotted as a function of the CH$_3$ photofragment translational energy. }
\label{fig_CH3I_PFB}
\end{figure}

The photodissociation experiments were performed on thin films (1--10ML) of the selected fluorobenzene with 1ML of CH$_3$I adsorbed on top. Unless otherwise noted, the samples were oriented so that the surface normal was in the direction of the QMS so that the laser light was incident at 45$^{\circ}$ from the surface normal. Figure {\ref{fig_CH3I_PFB}} shows TOF spectra obtained using p- and s-polarized light from 1ML CH$_3$I adsorbed on 3ML pFBz. The TOF spectrum obtained using p-polarized light (red trace) displays two peaks, labelled I (at 43$\mu$s) and I* (at 53$\mu$s). These peaks are qualitatively similar to those observed in gas-phase photodissociation of CH$_3$I at this wavelength and their origin can be understood from the description for neutral photodissociation outlined in Section {\ref{CH3X_pdissn}}. The inset plot shows the same data plotted in terms of the CH$_3$ photofragment kinetic energy-- the `I' peak pathway centered at 1.75eV and the `I*' pathway peak centered at 1.15eV. These peaks are prominent for p-polarized light when collecting CH$_3$ photofragments in the surface normal direction when adsorbed CH$_3$I molecules have their C--I bond axis oriented in the surface normal direction, so that the parallel transition $X\rightarrow {^3Q_0}$ is accessible from a component of the p-polarized light. When s-polarized light is used (blue trace) the parallel transition is forbidden in this detection geometry and a much smaller CH$_3$ photofragment yield is observed. The contributions to the observed s-polarized TOF spectrum likely come from several sources. The smaller `I' pathway peak and absence of a distinct I* peak appears to be a consequence of the $X\rightarrow {^1Q_1}$ (perpendicular) transition that has a smaller oscillator strength than the $^3Q_0$ counterpart, and leads exclusively to the I pathway in the dissociation limit. It is also observed that there is a continuum of CH$_3$ photofragment energies (i.e. a broad background), in the ToF spectra which we interpret as primarily due to orientation effects for the adsorbed CH$_3$I, which would include antiferroelectric ordering or disorder. Such order effects can lead to CH$_3$ photofragments undergoing collisions as they depart the surface region, so that their kinetic energies are reduced and CH$_3$ photofragments that were initially aimed in other directions are detected in the surface normal direction. The initial photodissociation event for these inelastically scattered CH$_3$ might also have a contribution from the stronger $^3Q_0$ transition from molecules not oriented in the surface normal direction.

\begin{figure}[h]
\includegraphics[scale=0.65]{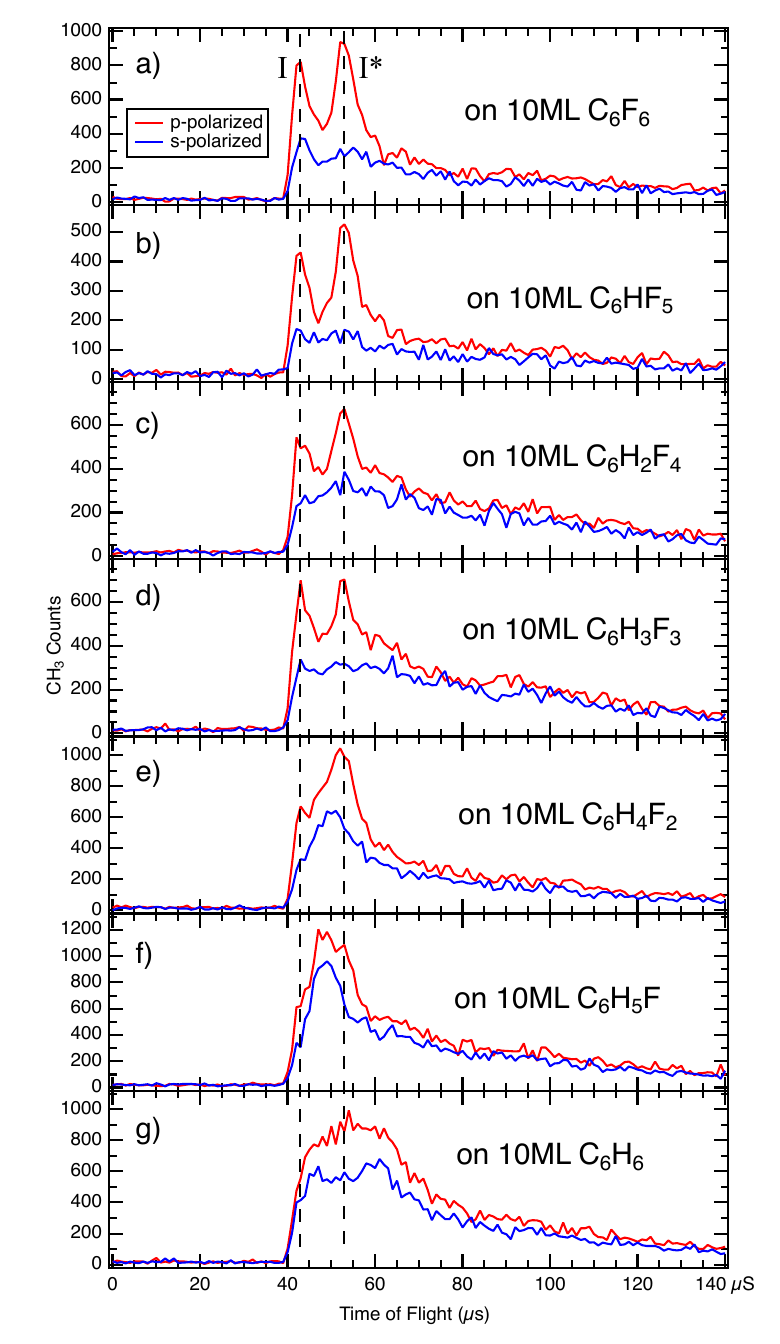}
\caption{Time-of-flight spectra for CH\(_3\) photofragments due to the photodissociation of 1ML CH\(_3\)I on 10ML thin films of (a) C\(_6\)F\(_6\) (hFBz); (b) C\(_6\)HF\(_5\) (pFBz); (c) 1,2,4,5-C\(_6\)H\(_2\)F\(_4\) (tetraFBz); (d) 1,3,5-C\(_6\)H\(_3\)F\(_3\) (triFBz); (e) 1,4-C\(_6\)H\(_4\)F\(_2\) (dFBz); (f) C\(_6\)H\(_5\)F (mFBz) and (g) C\(_6\)H\(_6\) (benzene), obtained using 248nm light. The red traces are obtained using p-polarized light, while the blue traces are obtained using s-polarized light. The CH$_3$ photofragments are detected in the surface normal direction. The vertical dashed lines indicate the nominal flight times for the CH$_3$ photofragment peaks from the I and I* A-band photodissociation pathways as described in the text.}
\label{fig_comparison_TOF}
\end{figure}

Time-of-flight spectra for CH$_3$I adsorbed on 10ML thin films for the series of fluorobenzenes and benzene are shown in Fig. {\ref{fig_comparison_TOF}}. There are systematic changes in the TOF spectra for the sequence of fluorobenzene thin films, most clearly observed in the s-polarization spectra that show the emergence of additional TOF spectrum features for dFBz, mFBz and benzene, and also in the corresponding p-polarization spectra where the I and I* peaks become less well resolved due to this same new feature overlapping in these spectra. 

For CH$_3$I adsorbed on 10ML hFBz and pFBz (Fig. {\ref{fig_comparison_TOF}}(a) and (b)) the TOF features are broadly similar to that described above for Fig. {\ref{fig_CH3I_PFB}} with the p-polarized spectra (red traces) showing well-resolved peaks for the I and I* dissociation pathways and a small I-pathway peak in the s-polarization data (blue traces). The broad CH$_3$ distribution at longer flight times is somewhat larger for the 10ML C$_6$F$_6$ film, which we believe is due to increased orientational disorder for the CH$_3$I as compared to the spectra from the thinner pFBz films (e.g. Fig. {\ref{fig_CH3I_PFB}}) and also for thinner C$_6$F$_6$ films which we described previously\cite{jensen:2021}. Based on this and other recent work studying the photodissociation of CH$_3$Br and CH$_3$Cl on pFBz at 193nm\cite{jensen:unpub}, we ascribe the somewhat lower background to better orientational ordering for the dipolar CH$_3$I molecules on the 10ML pFBz films as compared to hFBz. 

The TOF spectra for CH$_3$I adsorbed on 10ML tetraFBz (Fig. \ref{fig_comparison_TOF}(c)) and triFBz (Fig. \ref{fig_comparison_TOF}(d)) are quite similar-- the I and I* photodissociation peaks are clearly visible in the TOF spectra from p-polarized light (red traces) but less prominent than for the cases discussed above. There is a relatively higher intensity low-energy tail of CH$_3$ photofragment counts ($\tau$\textgreater 60$\mu$s) as compared to the other spectra in Fig. \ref{fig_comparison_TOF}. Similarly the TOF spectra for s-polarized light (blue traces) have higher relative CH$_3$ counts but in a broad TOF distribution without distinctive peaks. For the CH$_3$I on tetraFBz and triFBz thin films we conclude that the dissociation is dominated by neutral CH$_3$I photodissociation via the A-band mechanism described in Section {\ref{CH3X_pdissn}}, with orientational disorder for the CH$_3$I likely accounting for the higher CH$_3$ counts at long flight times. 

The TOF spectra for CH$_3$I adsorbed on dFBz, mFBz and benzene in Figs. {\ref{fig_comparison_TOF}}(e)-(g) differ substantially from the other spectra in the series. Using p-polarized light (red traces) the I and I* channel peaks are not well resolved as separate peaks, while the s-polarized light spectra (blue traces) show distinct features in the TOF spectra that are not observed for the other thin films. For dFBz (Fig. {\ref{fig_comparison_TOF}}(e)) the new feature appears to peak at roughly 51$\mu$s flight time, which is intermediate to that of the I and I* flight time peaks. For mFBz (Fig. {\ref{fig_comparison_TOF}}(f)) the new peak is at 49$\mu$s flight time. For benzene (Fig. {\ref{fig_comparison_TOF}}(g)), new peaks appear at 47$\mu$s and 60$\mu$s (peaks "A" and "B") with the first between the I and I* peaks and the second appearing at a later time\cite{jensen:2021}. In contrast to the I and I* peaks, which are most prominent when p-polarized light is used, these additional peaks are of similar intensity for both p- and s-polarized incident light. The TOF spectra obtained using p-polarized light are composed of both the neutral CH$_3$I photodissociation channel (I and I*) features, plus the new dissociation channel that has an intensity and peak location that substantially overlaps with the I and I* features. 

\begin{figure}[h]
\includegraphics[scale=0.65]{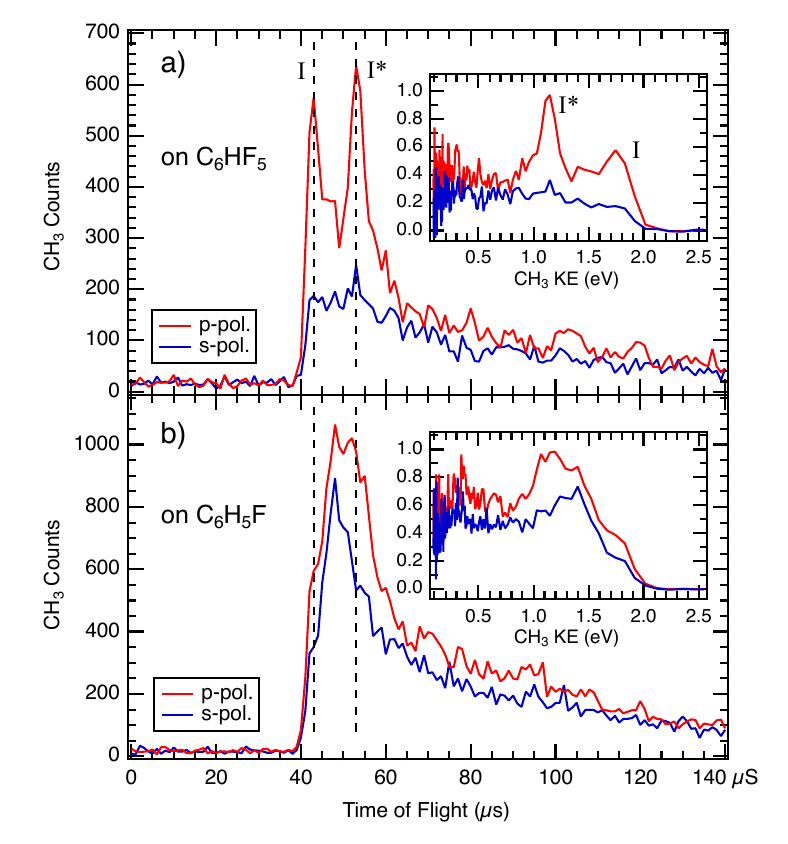}
\caption{Time of flight spectra for CH\(_3\) photofragments due to the photodissociation of 1ML CH\(_3\)I on 10ML (a) C\(_6\)HF\(_5\) (pFBz) and (b) C\(_6\)H\(_5\)F (mFBz) thin films, obtained using 248nm light. The red traces are obtained using p-polarized light and the blue traces are obtained using s-polarized light. The vertical dashed lines indicate the nominal flight times for CH$_3$ photofragment peaks from the I and I* A-band photodissociation pathways. The inset plots show the corresponding CH$_3$ kinetic energy distributions. The CH$_3$ photofragments are detected in the surface normal direction. }
\label{Fig_PFB_MFB_Comparison}
\end{figure}

More detailed comparison of TOF spectra from CH$_3$I photodissociation on 10ML films of pFBz and mFBz are shown in Fig. {\ref{Fig_PFB_MFB_Comparison}}. Using p-polarized light (red traces) the I and I* features are well separated for CH$_3$I on pFBz, while that for CH$_3$I on mFBz has strongly overlapped peaks. In comparing the p- and s-polarized TOF spectra, one can see that the I and I* pathways are present in the p-pol spectrum for CH$_3$I on mFBz with the additional intermediate TOF feature that overlaps the I and I* peaks. The inset figures show the same TOF data in terms of the CH$_3$ photofragment kinetic energies-- the I and I* peaks appear at 1.75eV and 1.15eV respectively, with the intermediate peak in centered near 1.40eV. In the TOF spectra we also observe that the total yield (sum of CH$_3$ counts above background) is substantially larger when using the mFBz thin film as compared to the pFBz thin film. A robust measure of this are the depletion cross sections obtained by the method described in Section {\ref{expt_detail}}. In comparing the depletion cross sections for the various thin films, we used CH$_3$I on the 10ML hFBz thin film as a reference so that the relative cross sections were compared to this system. To convert the cross sections to absolute values, the measured reference cross section for CH$_3$I on hFBz is $\sim\!8.0\times 10^{-19}$cm$^2$, similar to what was reported using 3ML hFBz films\cite{jensen:2021}. The relative depletion cross sections for CH$_3$I on the various 10ML thin films are shown in Table {\ref{enhancement_table}}. In comparison to the reference hFBz thin film, that from pFBz is very similar. In contrast, the relative depletion rate for CH$_3$I on the 10ML mFBz thin film is about 3 times larger. The larger depletion rate observed for CH$_3$I on the 10ML mFBz thin film and the features observed in TOF spectra for p- and s-polarized incident light support the proposal that a new dissociation pathway is present for the CH$_3$I adsorbed on top of the mFBz thin film.

\begin{table}[]
\begin{tabular}{|ccccccc|}
\hline
\multicolumn{1}{|c|}{hFBz} & \multicolumn{1}{c|}{pFBz} & \multicolumn{1}{c|}{tetraFBz} & \multicolumn{1}{c|}{triFBz} & \multicolumn{1}{c|}{dFBz} & \multicolumn{1}{c|}{mFBz} & \multicolumn{1}{c|}{C$_6$H$_6$} \\ \hline
\multicolumn{1}{|c|}{1}    & \multicolumn{1}{c|}{0.9}     & \multicolumn{1}{c|}{1.0}      & \multicolumn{1}{c|}{1.1}      & \multicolumn{1}{c|}{1.8}      & \multicolumn{1}{c|}{3.0}     & \multicolumn{1}{c|}{2.0}    \\ \hline
\end{tabular}

\caption{Relative depletion cross sections for 1ML CH$_3$I adsorbed on 10ML thin films of the selected fluorobenzenes and benzene using 248nm light. The data were normalized to the cross section observed from the system with hFBz (C$_6$F$_6$). }
\label{enhancement_table}

\end{table}

\begin{figure}[h]
\includegraphics[scale=0.65]{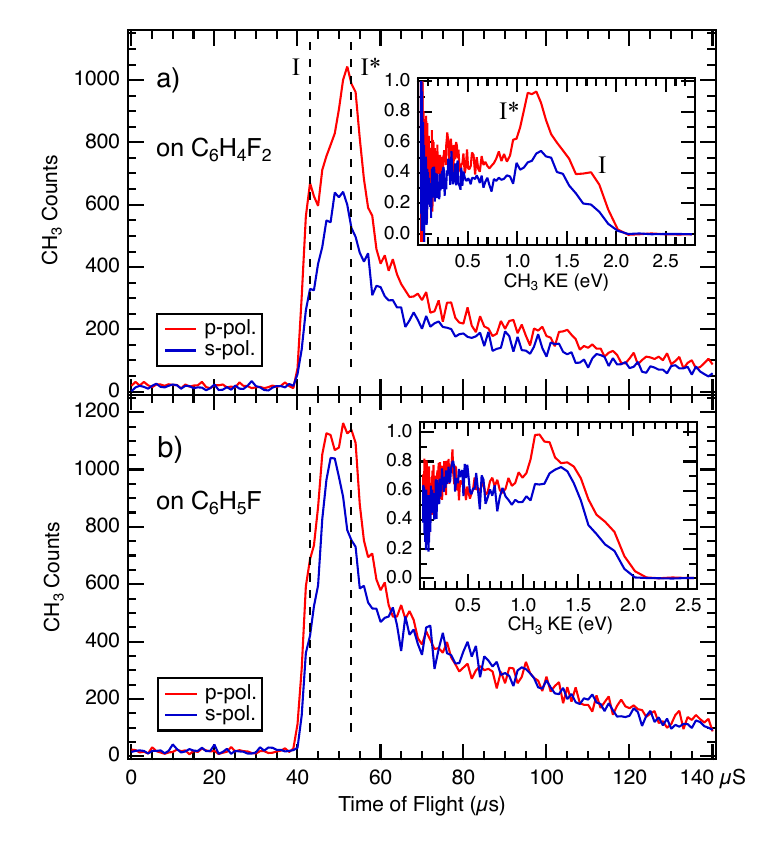}
\caption{Time-of-flight spectra for CH\(_3\) photofragments due to the photodissociation of 1ML CH\(_3\)I on 10ML (a) 1,4-C\(_6\)H\(_4\)F\(_2\) (dFBz) and (b) C\(_6\)H\(_5\)F (mFBz) thin films, obtained using 248nm light. The red traces are obtained using p-polarized light and the blue traces are obtained using s-polarized light. The vertical dashed lines indicate the nominal flight times for CH$_3$ photofragment peaks from the I and I* A-band photodissociation pathways. The inset plots show the corresponding CH$_3$ kinetic energy distributions. The CH$_3$ photofragments are detected in the surface normal direction.}
\label{Fig_DFB_MFB_Comparison}
\end{figure}

Figure {\ref{Fig_DFB_MFB_Comparison}} shows TOF spectra and photofragment kinetic energy distributions for CH$_3$I adsorbed on 10ML dFBz and mFBz. For CH$_3$I on the dFBz film, the s-polarization data (blue traces) show the peak to be somewhat lower intensity and a bit slower (51$\mu$s or 1.3eV) as compared to that for the mFBz thin film. The p-polarized TOF spectrum shows the I and I* neutral photodissociation pathways plus the new pathway that is observed for both polarizations. The relative CH$_3$I depletion cross section measured (Table {\ref{enhancement_table}}) for the dFBz thin film was enhanced by a factor of roughly 1.8 compared to the hFBz thin film. These observations support the notion that there is also an enhanced dissociation mechanism operating for CH$_3$I on 10ML dFBz thin films that is mediated by photoabsorption in the dFBz.

\begin{figure}[h]
\includegraphics[scale=0.65]{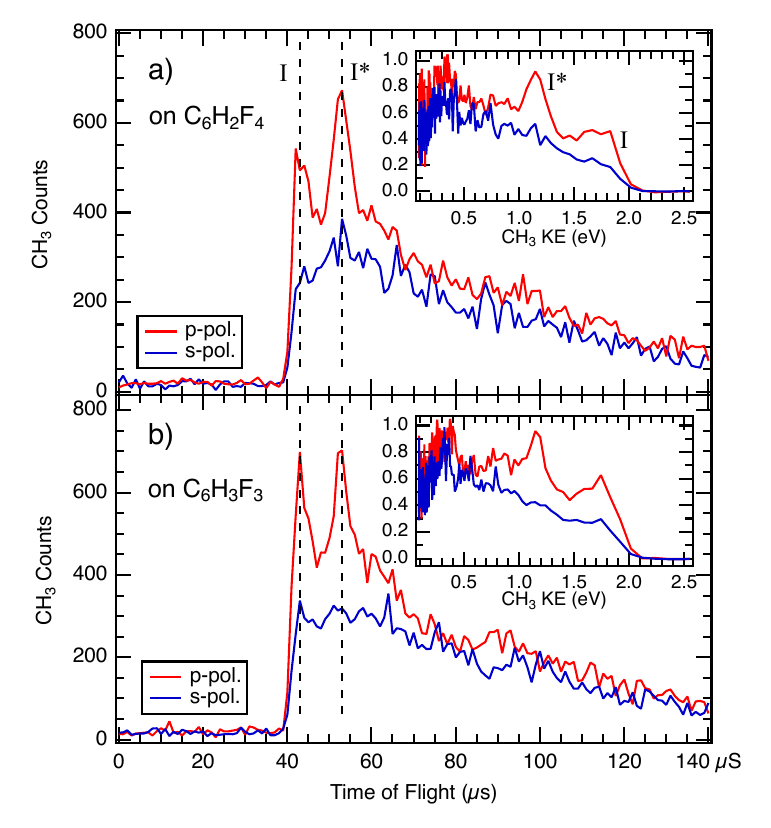}
\caption{Time-of-flight spectra for CH\(_3\) photofragments due to the photodissociation of 1ML CH\(_3\)I on 10ML (a) 1,2,4,5-C\(_6\)H\(_2\)F\(_4\) (tetraFBz) and (b) 1,3,5-C\(_6\)H\(_3\)F\(_3\)(triFBz) thin films, obtained using 248nm light. The red traces are obtained using p-polarized light and the blue traces are obtained using s-polarized light. The vertical dashed lines indicate the nominal flight times for CH$_3$ photofragment peaks from the I and I* A-band photodissociation pathways. The inset plots show the corresponding CH$_3$ kinetic energy distributions. The CH$_3$ photofragments are detected in the surface normal direction.}
\label{Fig_triFB_tetraFB_Comparison}
\end{figure}

For CH$_3$I adsorbed on 10ML tetraFBz and triFBz thin films, TOF spectra are shown in Fig. {\ref{Fig_triFB_tetraFB_Comparison}}. As noted above, the CH$_3$ photofragments in these thin film systems show a higher intensity broad background-- in the p-polarization light data the I and I* peaks are clearly visible and well-separated while the s-polarization data is almost featureless aside from small peaks that are present at the I and I* peak positions. The TOF data are not definitive in identifying any new dissociation pathways for CH$_3$I, such as seen for the mFBz and dFBz thin films. Measurements of the depletion cross sections for CH$_3$I on the tetraFBz and triFBz thin film systems in Table {\ref{enhancement_table}} also do not show a significant change in depletion rates as compared to that of the reference hFBz thin film, although the triFBz has a slightly larger rate ($\sim$10\%) which is small enough that if there is a contribution from a new dissociation pathway, we are unable to resolve it in the TOF data, given the broad spectra as seen in Fig. {\ref{Fig_triFB_tetraFB_Comparison}}.

\begin{figure}[h]
\includegraphics[scale=0.80]{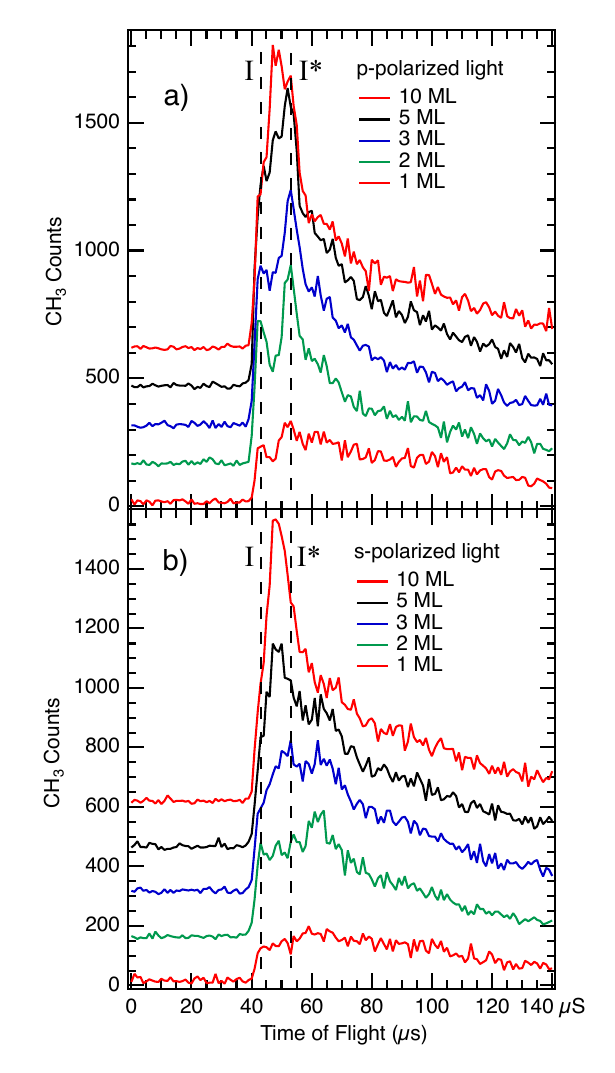}
\caption{Time-of-flight spectra for CH\(_3\) photofragments following the photodissociation of 1ML CH\(_3\)I adsorbed on varying thickness C$_6$H$_5$F (mFBz) thin films, obtained using 248nm light. The spectra in (a) are obtained using p-polarized light and the spectra in (b) are obtained using s-polarized light. Successive spectra are offset vertically by 150 counts for clarity. The CH$_3$ photofragments are detected in the surface normal direction. The vertical dashed lines indicate the nominal flight times for CH$_3$ photofragment peaks from the I and I* A-band photodissociation pathways.}
\label{Fig_varied_MFB_Comparison}
\end{figure}

The novel photodissociation pathways observed for CH$_3$I adsorbed on dFBz, mFBz and benzene thin films are only observed for multilayer films. Figure {\ref{Fig_varied_MFB_Comparison}} shows TOF spectra obtained using p- and s-polarized light for 1ML CH$_3$I on varying thickness (1--10ML) films of mFBz on Cu(100). As was observed for CH$_3$I on benzene\cite{jensen:2021}, CH$_3$ photofragments from CH$_3$I A-band photodissociation pathways are observed for all the films from 1ML through to 10ML. In the p-polarization TOF data, the I and I* pathways are observed distinctly for the thinner films (1--3ML), and as the film thickness is further increased (5 and 10ML) the new dissociation feature appears and grows in intensity such that the three peaks overlap. In the s-polarization TOF spectra, the new dissociation feature is clearly resolved and has high intensity for the 5 and 10ML thin films, possibly visible as a shoulder for the 3ML film in Fig. {\ref{Fig_varied_MFB_Comparison}}(b). These observations suggest that the new dissociation mechanism found for mFBz and dFBz films require a separation of the photoabsorbing molecules from the Cu(100) substrate in order to cause the dissociation of the coadsorbed CH$_3$I. In considering all of the features observed for this process, we conclude that the photochemical process occurs by photoabsorption in the underlying fluorobenzene thin film with subsequent electronic energy transfer to the CH$_3$I at the vacuum interface.

\begin{figure}[h]
\includegraphics[scale=0.73]{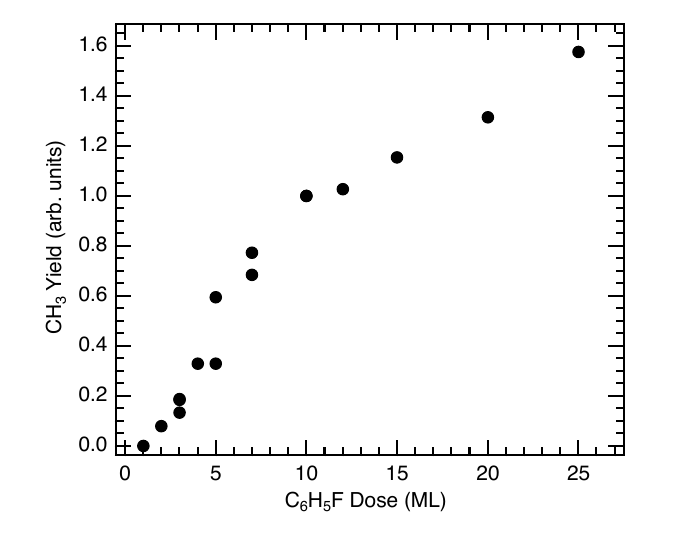}
\caption{Yield of CH$_3$ photofragments in the EET peak region from CH$_3$I on C$_6$H$_5$F (mFBz) films created from varying the dose. The yields were obtained from TOF spectra using s-polarized light. }
\label{Fig_CH3_yield_vs_dose}
\end{figure}

The growth of the photodissociation yield with the film thickness suggests that the initial photoabsorption is followed by energy transport, so leads to the question of what range is possible for this transport process before quenching occurs\cite{gebauer:2004}. The data of Fig. {\ref{Fig_varied_MFB_Comparison}} substantiate different timescales for the quenching of neutral photodissociation compared to the EET mechanism. For example, the p-polarization spectrum for CH$_3$I on 2ML mFBz clearly shows the I/I* pathways similar to Fig. {\ref{fig_CH3I_PFB}} with little or no observable contribution from the EET dissociation pathway, but as the mFBz film thickness is increased, the contribution from the EET pathway grows and eventually dominates the p- and s-polarized TOF spectra. At short separations, the quenching rate depends sensitively on wavefunction overlap between the excited state and substrate valence states. For the thinnest films (a few ML) resonant energy transfer in which electrons in the metal tunnel to fill the adsorbate hole state and the excited electron tunnels to unoccupied metal states\cite{gebauer:2004,Zhou:1995,avouris:1984} is the most likely quenching mechanism. The timescale for the dissociation of the CH$_3$I moiety on purely repulsive potential energy surfaces (e.g. $^3Q_0$) will be roughly the same independent of mechanism. This suggests that the characteristic timescale for the thin film excitation is significantly longer for the EET-driven dissociation than the neutral photodissociation, which occurs in $\sim$100\,fs\cite{Eppink:1998, alekseyev:2007}. A recent study of luminescence from exciton states in organic thin films grown on single crystal metal substrates found that lifetimes (varying from picosecond to nanosecond timescales) increased by an order of magnitude for films $\sim$10ML thick as compared to those few monolayers thick\cite{stallberg:2019}. In order to focus on the EET mechanism and minimize the contribution from CH$_3$I neutral photodissociation, the CH$_3$ yield in the EET peak (centered around 48$\mu$s flight time) from CH$_3$I on mFBz using s-polarized light was used (similar to the spectra in Fig. {\ref{Fig_varied_MFB_Comparison}}b). The CH$_3$ EET photoyield shown in Fig. {\ref{Fig_CH3_yield_vs_dose}} is found to increase roughly linearly from 1ML up to $\sim$10ML doses, and continue to increase with reduced slope ($\sim$1/3 that at lower doses) up to the thickest films used (25ML doses). That the observed yield curve does not saturate within the range of doses studied could be due to further reduction of longer range dipole quenching\cite{chance:1975}, but it is more likely due to the morphology of the thin films grown under the conditions of our experiments.

\subsection{Effects of Fluorobenzene Film Morphology}
\label{filmmorph}
\begin{figure}[h]
\includegraphics[scale=0.58]{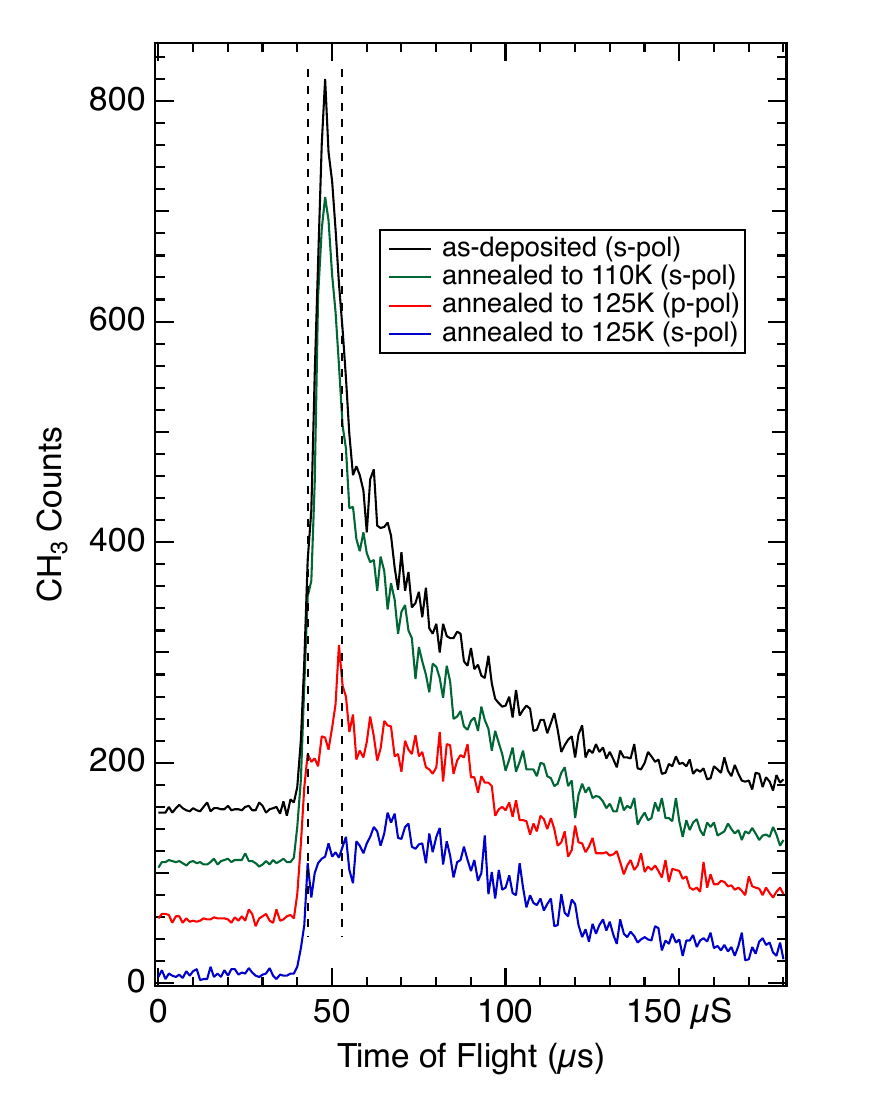}
\caption{Time-of-flight spectra for CH\(_3\) photofragments from the photodissociation of 1ML CH\(_3\)I adsorbed on 10ML  C$_6$H$_5$F (mFBz) thin films as-deposited (at T\textless 100K) or that have subsequently been annealed and re-cooled to T\textless 100K. The vertical dashed lines indicate the nominal flight times for CH$_3$ photofragment peaks from the I and I* A-band photodissociation pathways. The spectra are offset vertically by 50 counts for clarity.}
\label{Fig_warming_mFBz_ToF}
\end{figure}

The increasing CH$_3$ yield observed for CH$_3$I on 1--10ML mFBz films is due in part to the reduced quenching of excitations as the distance to the Cu(100) metal surface increases for the reacting species. For thicker films, quenching is less efficient so that the excitation can be formed and have sufficient lifetime and mobility to diffuse and cause dissociation of CH$_3$I by EET from mFBz. A complication is that the effect of the thin film morphology appears to become more significant as the mFBz dose increases. The most satisfactory explanation for the observed behaviour at higher doses in Fig. {\ref{Fig_CH3_yield_vs_dose}} is that the mFBz thin films have 3D island crystallites forming at higher coverages, analogous to the behaviour found for benzene multilayer films grown under similar conditions\cite{jakob:1996}. In the present experiments, we have grown the mFBz (and the other fluorobenzene) thin films at substrate temperatures below 100K, so that these films are crystalline (i.e. grown above temperatures where amorphous films would be expected, \textless 60K). The morphology of benzene multilayer films grown in similar conditions has been studied in some detail\cite{lee:2006, xi:1994,jakob:1996, jakob:1989}, but there is little in the current literature specific to fluorobenzene film growth. It is quite likely that the first layer on Cu(100) fully covers the metal surface, given the observations from TPD described in Section {\ref{TPD_section}}. For higher doses, the structure of the thin films is less clear, but most likely the Stranski-Krastanov type where crystallites grow on the surface so that regions have differing thickness, dictated in part by the mobility of the molecules as they arrive at the surface. We have done a limited study of the effect of annealing the mFBz thin films. Figure {\ref{Fig_warming_mFBz_ToF}} shows the effect of annealing on the TOF spectra from mFBz thin films. The experimental procedure used was to dose the mFBz thin film (10ML) at T\textless 100K, then warm and maintain the substrate at a target temperature for a minute, and re-cool to T\textless 100K followed by dosing the CH$_3$I layer. Using the TOF spectra as a probe, we found little change until annealing temperatures of 120K or higher were used. The highest annealing temperature used was 130K to prevent mFBz desorption, which begins near 150K. The TOF spectra from mFBz thin films annealed to 120K\textless T\textless 130K were significantly altered from the as-deposited thin films-- the large signal from EET causing CH$_3$I dissociation (peak at 48$\mu$s) was effectively absent (s-polarization) as were the I/I* pathway peaks characteristic direct photodissociation of CH$_3$I (p-polarization). The effect of annealing the mFBz at T\textgreater 120K appears to be to allow adsorbate mobility to grow larger crystallites of mFBz and leave large terraces of the Cu(100) substrate with low coverage (presumably monolayer mFBz) so that much of the subsequently dosed CH$_3$I is close enough to the metal surface for efficient quenching of the excitation-- similar to the low coverage (1ML) traces in Fig. {\ref{Fig_varied_MFB_Comparison}}. The TOF spectra from annealed films have CH$_3$ translational energy distributions very different from that seen for as-deposited 10ML mFBz films. Our findings suggest a similar conclusion to that made for benzene thin films: ``extra thermal annealing leads to a destabilization of smaller crystalline domains in favour of larger crystallites which then grow even more''\cite{jakob:1996}. Thus for the thin films grown at low temperature (T\textless 100K), we expect that these are crystalline and for multilayer doses, have varying thickness crystalline regions with an average thickness as indicated by the flux of molecules used. The primary evidence for this from Fig. {\ref{Fig_CH3_yield_vs_dose}} is that the CH$_3$ photoyield does not saturate for some mFBz dose, which would be expected for layer-by-layer growth\cite{gebauer:2004}. To the extent we have investigated, annealing these films leads to growth of a small number of larger crystallites that cover much less of the surface that leaves the majority of the substrate covered by monolayer quantities of the mFBz.

\section{Additional Discussion}
The observation of an EET dissociation mechanism for CH$_3$I adsorbed on the mFBz, dFBz and benzene thin films but absent for the higher fluorobenzenes is at first surprising as the photoabsorption cross sections for the higher fluorobenzenes are generally as large or larger than for benzene at the 248nm wavelength used in the present work\cite{dawes:2017, philis:1981}. However as the fluorine atom number increases the `perfluoro effect' lowers an unoccupied $\pi\sigma$* level toward the LUMO $\pi\pi$*. The lowered $\pi\sigma$* state increases the rate of internal conversion $S_1 \rightarrow S_0$, reducing the excited state lifetime and fluorescence yield significantly as compared to lower F-count fluorobenzenes\cite{zgierski:2005}. This effect seems to be responsible for the absence of the EET pathway being observed for CH$_3$I on the triFBz through hFBz thin films.

For CH$_3$I adsorbed on the pFBz and the other fluorobenzene thin films we found no evidence for dissociative electron attachment (DEA) of the CH$_3$I due to photoelectrons using 248nm light, which is consistent with the findings we previously reported for a series of halomethanes adsorbed on benzene and hFBz thin films\cite{jensen:2021}. We have observed features consistent with the photoelectron DEA mechanism for CH$_3$Br and CH$_3$Cl for very thin films of benzene and several of the fluorobenzenes (1-2ML films, using both 248nm and 193nm light)\cite{jensen:unpub}, but which diminishes as the film thicknesses are increased and was not observable for the thicker multilayer films. Further, 248nm illumination of CH$_3$Br on 10ML mFBz and dFBz thin films found no CH$_3$ photofragment signal in TOF spectra, which we interpret as being due to the analogous `A-band' dissociative states for CH$_3$Br being at higher energies\cite{kato:2010} than for CH$_3$I and not accessible via 248nm photoexcitation. 

The energies observed for the CH$_3$ photofragments from the EET process can be used to estimate the excitation energy causing dissociation (equivalent to $h\nu$ in Equ. {\ref{Equ_2}}). The excitation energies found are 4.71eV (C$_6$H$_6$), 4.60eV (mFBz) and 4.49eV (dFBz) if $E_{int}(CH_3)$ is assumed to be the same as seen in the neutral photodissociation (e.g. Fig. {\ref{fig_CH3I_PFB}} (top)) as well as the I-pathway outcome, so represent a rough upper bound on the excitation energy causing dissociation. These can be used as a basis for considering more specific detail for the mechanism of dissociation.

The evidence for the observed EET to CH$_3$I in the thin film points to an initial photoabsorption by the $S_1$ state of the aromatic molecules. In solid benzene at 100K the $S_1$ fluorescence lifetime is roughly 80ns\cite{lumb:1971}, and though no data exists for the mFBz or dFBz in the solid state, gas-phase $S_1$ fluorescence lifetimes are longer (10--20ns) than for the higher fluorobenzenes\cite{phillips:1972}. In EET systems the energy transfer between donor and acceptor generally occurs via states of the same multiplicity. For CH$_3$I the only available singlet state in the A-band is the $^1Q_1$, which leads to the I-pathway in Equ. \ref{Equ_1}. However in the Franck-Condon region, the preponderance of the $^1Q_1$ state is located at higher energy\cite{alekseyev:2007, kato:2010} than the available excitation energy, so seems unlikely to be the acceptor state for EET. For mFBz and dFBz, only a single TOF peak is observed, at lower CH$_3$ photofragment kinetic energy than the $^3Q_0$ neutral photodissociation I-pathway but higher energy than the I*-pathway. This suggests an excited state energy less than the initial $h\nu =4.99$eV photon that leads to the I-pathway only. There are several lower energy electronic excited states for CH$_3$I exclusively leading to the I-pathway that are derived from the $^3Q$ multiplet, including the $^3Q_1$ as well as others that are optically dark\cite{alekseyev:2007}. In contrast, the CH$_3$I on benzene system displays bimodal EET features in the TOF spectrum (peaks ``A''and ``B''), as discussed previously\cite{jensen:2021}, and was ascribed to EET via the CH$_3$I $^3Q_0$ excited state. These TOF peaks would correspond to the I and I*-pathways, and the higher initial excitation energy is consistent with the ordering of the $^3Q$ states in the Franck-Condon region. To access CH$_3$I triplet state excitations through EET, the heavy-atom effect\cite{Koziar:1978,martinho:2001} due to the presence of I-atoms would be required to facilitate the mixing of states of differing multiplicity and intersystem crossing, such as that seen in the $X\,^1\!A_1\rightarrow {^3Q_0}$ photoexcitation and in solution-phase fluorescence quenching of pyrene by CH$_3$I\cite{martinho:2001}.

For the arenes used in the present work, the lowest energy triplet states lies well below the $S_1$ states and are at too low an excitation energy to cause the observed CH$_3$ photofragment kinetic energies, but the second triplet states are located in the same energy region as the $S_1$ states. For solid benzene the triplet $T_1$ origin is at 3.67eV, and the $T_2$ origin is at 4.59eV ($S_1$ at 4.69eV)\cite{dawes:2017,swiderek:1996}. The corresponding states for the mFBz and dFBz states are at similar or slightly lower energies in the gas-phase\cite{phillips:1972} and thin films solids\cite{swiderek:1997}. 

The suggestion that the lower energy EET processes seen for CH$_3$I on mFBz and dFBz films leads to a different excited state than that seen on benzene requires some scrutiny-- for optical excitation throughout the A-band region, the $^3Q_0$ excitation dominates. For the listed $^3Q$ states to be significant in the EET process, the wavefunction overlap with arene excited states must be the critical factor, which will depend on the state energies, adsorption site and geometry for CH$_3$I on the different molecular thin films. 

It is useful to briefly consider comparisons with some findings from the photochemistry of bimolecular clusters, though these represent the converse from the extended condensed-phase systems of the present work. The benzene:I$_2$ dimer (as well as other arene:I$_2$ dimers) has been studied extensively due to the formation of a well-known charge-transfer (CT) complex. Photodissociation dynamics following excitation of the CT state find that the ionic and neutral state mixing leads to a significant dissociation pathway C$_6$H$_6$-I + I with substantial kinetic energy release. Modelling of this process finds that the pathways and outcome are sensitive to the initial geometry of the dimer\cite{cheng:1996,deboer:1996}. Another cluster system of relevance is that of the CH$_3$I:O$_2$ dimer, where near-UV photodissociation led to CH$_3$I dissociation ascribed to the $^3Q_1$ state, as the I-pathway outcome of Equ. {\ref{Equ_1}} was the only one observed\cite{bogomolov:2016}, similar to the observations and argument made above for CH$_3$I on mFBz and dFBz thin films. 

\section{Summary and Conclusions}
The 248nm illumination of CH$_3$I adsorbed on various fluorobenzene films grown on a Cu(100) substrate displays an additional CH$_3$I photodissociation mechanism for 1,4-difluorobenzene, monofluorobenzene and benzene thin films but absent for the higher F-count fluorobenzenes studied. The use of s-polarized incident light allows this mechanism to be examined separately from gas-phase-like neutral photodissociation of CH$_3$I. This novel photodissociation mechanism displays a unimodal CH$_3$ photofragment kinetic energy distribution for the fluorobenzenes, with peak kinetic energies that are distinct for each molecular thin film. The substrate-induced quenching behaviour observed suggests that the excitation has a notably longer lifetime than that for the neutral photodissociation channel. These observations can be understood by a model in which the photoexcitation creates excitons in the fluorobenzene thin films that transfer this energy to the CH$_3$I, which promptly dissociates. As such, we anticipate that this phenomenon should be operative for a wide variety of aromatic molecules that display similar photoexcitations in the near-UV region.


\section*{References}
\bibliography{CH3I_Fluorobenzenes_EET_Bibliography}

\end{document}